\documentclass[aip,jcp,graphicx]{revtex4-1}
\usepackage{graphicx}% Include figure files

\begin{document}

%\preprint{}

\title{Pimples reduce and dimples enhance flat dielectric surface image repulsion.}
% Force line breaks with \\

\author{Francisco J. Solis}

\email{francisco.solis@asu.edu}

\affiliation{School of Mathematical and Natural Sciences, Arizona State University, Glendale, Arizona,  85306}

\author{Monica Olvera de la Cruz}%

\affiliation{Department of Chemical and Biological Engineering, Northwestern University, Evanston, IL 60208, United States}

\affiliation{Department of Physics and Astronomy, Northwestern University, Evanston, IL 60208, United States}

\affiliation{Department of Materials Science and Engineering, Northwestern University, Evanston, IL 60208, United States}

\date{\today}% It is always \today, today,

\begin{abstract}
Near solid-liquid or liquid-liquid interfaces with dielectric contrast,
charged particles interact with the induced polarization charge
of the interface. These interactions contribute to an effective self-energy of the bulk ions and mediate ion-ion interactions. For flat interfaces, the self-energy and the mediated interaction are neatly constructed by the image charge method. For other geometries,
explicit results are scarce and the problem must be treated via approximations
or direct computation. This article provides analytical results, valid
to first order in perturbation theory, for the self-energy of particles
near a deformed near-flat interface. Explicit formulas are provided for
the case of a sinusoidal deformation; generic deformations can then be treated by superposition.
In addition to results for the self-energy,
the surface polarization charge due to a single ion is presented as a quadrature. The
interaction between an ion and the deformed surface is modified by
the change in relative distance as well as by the curvature of the
surface. Solid walls with a lower dielectric constant than the liquid
repel all ions. We show here, however, that the repulsion
is reduced by local convexity and enhanced by concavity; dimples are
more repulsive than pimples. 
\end{abstract}


\maketitle %\maketitle must follow title, authors, abstract and \pacs


\section{Introduction}\label{section:introdcution}

The properties of electrolyte solutions can be greatly modified by
their interaction with their confining surfaces, particularly in the
presence of a mismatch between the properties of the bulk and the
bounding material. This is the case for confined electrolytes \citep{Wagner1924,Onsager1934}, and more specifically those with graphene or silica substrates \citep{Jurado2017}, in capacitive systems \cite{Simon2008}, in colloidal suspsensions \citep{Sacanna2007}, polyelectrolyte solutions\citep{Shim2012249}, encapsulated or disordered proteins \citep{Cao2017600,Nott2016569}, and liquid-air interfaces \cite{Levin2009} among others. 
Features of these systems remain a continuous focus of fundamental and applied research
by means of both theoretical and experimental methods as well as investigations
in their technological deployment. Possible applications where these properties are important include processes of colloidal stabilization\citep{Vrijenhoek2001,Hansen2000},
control of surface tension \citep{Levin2009},
design of electrochemical capacitors\citep{Simon2008}, and the use
of membranes as ion transfer methods\citep{SHIRAI1995,DILGER1979}. 

Complex behavior can arise when the interface geometry
is modified. Most analytical and numerical investigations
assume interfaces with planar or other simple geometries \cite{Allen20014177,Messina2002, Santos2011,jadhao2012simulation,Arnold20134569,Barros2014,Gan2015,Shen2017, wu2018asymmetric,nguyen2019incorporating}. It has been
shown, however, that the roughness of the interface can have important
effects: forces between approaching surfaces are modified\citep{Bhattacharjee1998,Bowen2000,Bowen2002,Bradford2013}
and ion distributions can exhibit coupling to a substrate geometry\citep{wu2018asymmetric}. 

This article provides an evaluation of the contribution to the self-energy
of bulk ions due to their interaction with a structured polarizable
interface. An ion near the interface creates an induced surface polarization
charge that interacts with the ion itself, giving rise to this self-interaction.
For a flat surface, this self-energy can be readily described by the
image charge method. In the general case, it can be determined by solving 
an integral equation. In this investigation, the surface is considered
as a deformed plane. The surface polarization and the resulting self-energy
are determined to first order in the amplitude of the plane deformations. It is simplest to consider sinusoidal deformations and
construct solutions for all other shapes by superposition. The self-energy
in the presence of a sinusoidal deformation is analytically expressed,
while the surface charge density is presented as a quadrature that
can be directly evaluated. 

Analysis of electrolyte systems has multiple aspects. It requires
consideration of  the interaction between ions and the ensuing collective
behavior, the steric effects that arise from the presence of a solid
wall and the finite size of ions, and, as noted, the electrical interaction
between the ions and the interface as well as the ion-ion interaction
due to the polarization of the interface. This work is a contribution
to the understanding of this third component. The role of the interaction
with the dielectric discontinuity is surprisingly rich and challenging.
It was first studied by Wagner and Onsager\citep{Wagner1924,Onsager1934}. In that work the effects of the surface were described using the image method
and it was established that there is always a strong interaction between
ions and surface. In more recent work, Levin\citep{Levin2009} has analyzed the effect on the surface tension of a liquid-gas interface due to ions localized at the interface. These effects are
not immediately captured by mean-field theories, which require modification
to include self-energies \citep{netz2003variational,buyukdagli2010variational,wang2010fluctuation,wang2013effects}.
These examples show that a clear description of the properties of
the self-energy of ions near surfaces is a crucial element of progress
towards the understanding of these rich and complex systems. 

The results of this calculation presented here fulfill several
goals. The concrete evaluation of the self-energy provides
a direct way to assess the relevance of these effects in concrete situations.
As the results are analytical they can be used as a gauge to measure the
effectiveness of numerical solutions. They provide 
a way to incorporate roughness into effective theoretical descriptions
of the system and simplified numerical schemes. By identifying
the source of the modifications of the self-energy, the results produce heuristic
tools for the design of functional structures where variations of local ion populations are important. 

The explicit determination of the effects of the surface deformation requires considerable technical steps. It is, therefore, useful to summarize here some central findings.  
The key result of this analysis is that near a deformed surface, an ion of charge $ve$  acquires an excess self-energy $\Delta U$.  This modifies the classical self-energy result $U_{self}\sim q^2/z$ when the charge is at a distance $z$ from the surface, as obtained by the classical image charge method. A sinusoidal roughness of the form $h=A \cos (kx+\psi)$, with $x$ running along the surface and $\psi$ a relative phase, produces an excess self-energy described in detail in section \ref{section:results}. This explicit form is best understood by looking at its decay away from the surface and at its values very close to the deformation. For the first case, up to some constant numeric factors, an  approximate version of the result can be written as 
\begin{equation}
    \frac{\Delta U}{k_B T} \approx \frac{\ell_B A k^2}{(k z)^{3/2}} \cos(kx +\psi) \exp(-kz) [4\gamma -\gamma^2 (kz)].
\end{equation}
Here $\ell_B=e^2/(4\pi\varepsilon_W\varepsilon_0 k_BT)$ is the Bjerrum length, $k_bT$ the Boltzmann factor, $\varepsilon_0$ the vacuum permittivity and $\varepsilon_W$ the relative permittivity of the liquid phase. The ratio $\gamma=\Delta\varepsilon/\varepsilon_I$ is a factor that indicates the dielectric contrast between the liquid region and substrate, $\Delta\varepsilon$ is the difference in permittivities and $\varepsilon_I$ their average. In part, the result is as expected; effects of the roughness follow the pattern of the substrate and decay into the bulk with a characteristic length set by the roughness, namely, the wavelength $\lambda=2\pi/k$. The distance power $z^{-3/2}$ makes the decay region only important near the surface so that for large wavelengths the corrections do not penetrate indefinitely into the liquid bulk. There is, however, a surprising element in the result as well. The two terms in the expression, proportional to $\gamma$  and $\gamma^2$, have different signs and functional behavior. The first term reflects, roughly, the direct effect of the topography bringing the surface closer or farther away from the ion. The second is, in essence, proportional to the local curvature of the surface. When ions are very close to the surface, this second term creates en effective reduction in repulsion for ions near the convex crest of the rough substrate while enhancing the repulsion at the concave troughs.  These results suggest interesting applications of substrate patterning to enhance or deplete ion populations at specific locations.

The rest of the article is structured as follows. Section \ref{section:polarization} establishes
some basic notation and, before any direct calculation, describes
the origin of the results that are explicitly determined later on.
Section \ref{section:integral} describes the integral equation for the polarization charge
and describes the self-energy computed from interaction with this
charge. Section \ref{section:perturbation} develops the perturbative scheme used for the calculation.
Key results of the calculation are presented in section \ref{section:results}; examples
of evaluation of self-energies in specific geometries are provided
there as well. Section \ref{section:near} looks in more detail at the results when
the point of observation is very close to the surface. Some further conclusions are presented in \ref{section:conclusions}. Details of the evaluation of several expressions are presented in the appendices. 

\section{Geometric effects on polarization. }

The role of substrate roughness has been investigated in the context
of surface-surface interactions providing modified effective potentials that correct calculations for perfectly flat surfaces \citep{Bhattacharjee1998,Bowen2000,Bowen2002,Bradford2013}.
The origin of this behavior is easy
to understand as the roughness locally modifies the distance between corresponding patches of the interacting surfaces. In the case of where electrolyte ions
interact with surfaces a similar issue arises, though the results
are a more complicated. Before describing in detail the calculation
of the ions' effective self-energy, it is convenient to discuss the key effects that appear in the result. 

\begin{figure*}
\includegraphics{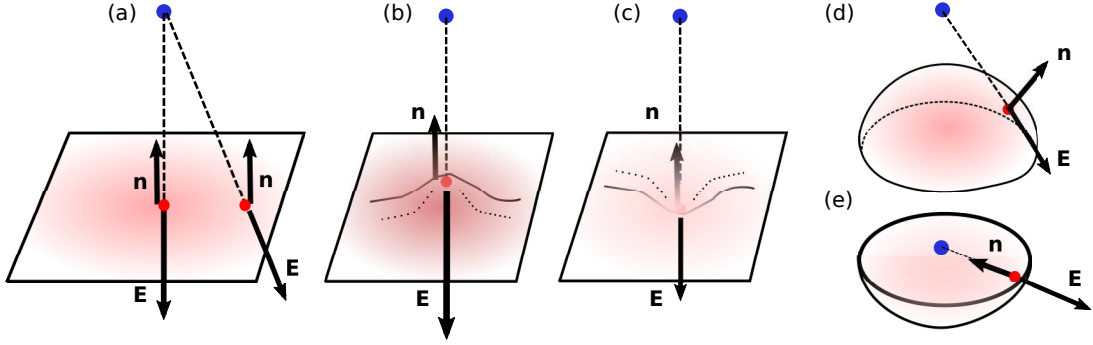}
\caption{\label{fig:Topography}Effects of topographical
deformations. (a) The induced surface charge density generated by
a single ion is largest at the location closest to the ion. At this
point the field is strongest and the normal is aligned with the field
direction. (b) At the crest of a deformation, a pimple, the distance
is smaller and the field and charge density is larger. (c) At the
bottom of a dimple, the field and charge density are smaller. (d) Convexity reduces the polarization misalignment of the normal and the field. (f) Concavity
increases the polarization enhancing the alignment. }
\end{figure*}

The interaction between a single ion and a surface arises from the
polarization of the interface between the liquid medium containing
the ion and its bounding solid material. The polarization charge appears
as the dielectric constants of both media are different. The net electric
field $\mathbf{E}$ produced by an ion and the induced surface charges create
a polarization field $\mathbf{P}=\varepsilon_{0}\chi\mathbf{E}$ where
$\chi$ is the polarizability. The polarization field 
produces a net surface charge density $\mathbf{\sigma=P}\cdot\mathbf{n}$,
with $\mathbf{n}$ the normal to the surface, pointing away from the
polarized medium. There is a contribution from each of the media and,
as they have different polarization magnitudes, the net surface charge
is not zero. If $\alpha$ labels the liquid region where the ion is
located and \textbf{$\beta$} labels the solid region, $\sigma=\chi^{(\alpha)}\mathbf{E^{(\alpha)}}\cdot\mathbf{n}^{(\alpha)}+\chi^{(\beta)}\mathbf{E^{(\beta)}\cdot n^{(\beta)}}=-(\chi^{(\alpha)}\mathbf{E^{(\alpha)}}-\chi^{(\beta)}\mathbf{E^{(\beta)}})\cdot\mathbf{n}^{(\beta)}.$
The field is discontinuous due to the very presence of charge, but
the contribution due to the ion is the same across the interface.
Therefore, just due to the field of the ion ($\mathbf{E}_i$ we have $\sigma=-\Delta\chi\mathbf{E}_i\cdot\mathbf{n}$,
with $\Delta\chi=\chi^{(\alpha)}-\chi^{(\beta)}$ and $\mathbf{n}$
the normal oriented into the liquid phase. Fig. \ref{fig:Topography} sketches the effects of geometry on this polarization contribution. In panel (a)
the induced polarization just
due to the ion field at a flat interface decreases away from the ion
due to the increased distance and the reduced alignment of the field
and the normal. 
Near a pimple, as in panel (b), the distance to the surface is smaller,
the field is stronger and the polarization is larger. The opposite
effect occurs for a dimple, as in panel (c). However, the distance
change is not the only effect at play. The field and the normal are
further misaligned in the neighborhood of a convex region, as shown
in (d), while their alignment is enhanced near concavities as shown
in (e). 
%
% \begin{figure}
% \end{figure}

It is important to remark that for a fixed point in the bulk the presence of a pimple creates both a polarization 
a enhancement, due to reduced distances, and a decrease due
to its convexity; opposite effects arise for the dimple.
The explicit calculation below shows that both effects are important.
When points with constant height about the average flat plane are
considered, the results show a dominant effect due to the decreased
or increased local height. However, if one samples the positions above
the surface maintaining the local distance to the surface constant,
the curvature effect becomes dominant. 

The effects noted above have considered only the effect of the direct field created by the ion. The calculation below contains, in addition, all effects due to interaction between the induced charges. The final results show that the heuristics above survive the inclusion of these further interactions. Namely, we still observe that convexity reduces the strength of the interaction while convexity enhances it. 

\section{Integral equation and self-energy}\label{section:integral}

The integral equation for the polarization is well known and is the basis of boundary methods for evaluation of ion interactions in the presence of dielectric media \citep{Allen20014177, Arnold20134569,jadhao2012simulation,Gan2015}. Here, its
structure is reviewed and, along the way, notation is established
for the rest of this article. The electric potential $\phi$ created
by free charges with density $\rho$ in a polarizable medium is given
by the solution of the Poisson equation:
\begin{equation}
\nabla\cdot\varepsilon_{0}\varepsilon\nabla\phi=\rho.\label{eq:PoissonDiff}
\end{equation}
The polarization vector is $\mathbf{P}=\varepsilon_{0}\chi\mathbf{E}=-\varepsilon_{0}\chi\nabla\phi$,
so that the volumetric polarization charge density is associated polarization
charge is
\begin{equation}
\omega_{v}=\nabla\cdot\mathbf{P}=-\varepsilon_{0}\nabla\cdot\chi\nabla\phi.\label{eq:PolarizationChargeDef}
\end{equation}
The electrostatic energy of the system can be evaluated from the expression 
\begin{equation}
U=\frac{1}{2\varepsilon_{0}}\int dV_{r}\int dV_{r'}\:\rho(\mathbf{r})G_{r,r'}[\rho(\mathbf{r}')+\omega_{v}(\mathbf{r}')].\label{eq:EnergyPolarization}
\end{equation}
Here the geometric Green function is $G(\mathbf{r},\mathbf{r}')=[4\pi|\mathbf{r}-\mathbf{r}'|]^{-1}$.
Inside the integral this is written as $G_{r,r'}$. If the charges
are considered point-like, their direct self-energy must be subtracted,
tough their self-interaction via polarization charges must still be
considered and is finite. 

The Poisson equation along with the definition of the polarization
charge implies an integral equation relation that is useful for explicit
computations. This is 
\begin{equation}
\omega_{v}(\mathbf{r})=\nabla_{r}\cdot\left[\chi(\mathbf{r})\nabla_{r}\int dV_{r'}\,G_{r,r'}\left(\omega_{v}(\mathbf{r}')+\rho(\mathbf{r}')\right)\right].\label{eq:IntegralFromPolarization}
\end{equation}
In the case where the system is composed of piece-wise regions of
uniform permittivity, the polarization charge appears only at the
location of the point charges, dressing them, and at the interfaces
between the uniform regions. For given point charges, this relation
becomes a boundary integral equation for the polarization charge. 

To make the presentation of the perturbative calculation more transparent
it is convenient to present the equation in a way that integral symbols
and coordinates are not explicit. A few preliminary steps are required.
First, it is assumed that the free charges are point-like, so that
the free charge can be written as $\rho(\mathbf{r})=\sum_{i}q_{i}\delta^{3}(\mathbf{r}-\mathbf{r}_{i})$.
At each of the locations of the free particles there is polarization
charge $\omega_{v}(\mathbf{r})=\omega_{i}\delta^{3}(\mathbf{r}-\mathbf{r}_{i})$
with $\omega_{i}=-\chi_{i}q_{i}/(1+\chi_{i})$ where $\chi_{i}$ is
the polarizability at the $i$-th location. The dressed charge at
each of these locations is the bare charge $q$ multiplied by the
factor $\varepsilon^{-1}.$ That is, $q_{i}+\omega_{i}=\varepsilon_{i}^{-1}q_{i}$.
The sum of the charge and its dressing polarization is $\rho(\mathbf{r})+\omega(\mathbf{r})=\sum_{i}\varepsilon_{i}^{-1}q_{i}\delta^{3}(\mathbf{r}-\mathbf{r}_{i})$. Next, the application of the gradient to the integral terms proportional to the surface charge results in two terms. One is proportional to the gradient of the polarization $\nabla \chi=\nabla\varepsilon$. The second is proportional to $\chi\int\nabla_r^2 G_{r,r'}\omega(\mathbf{r}')=-\chi\omega(\mathbf{r})$. This contribution can be collected on the left hand side to obtain $(1+\chi_I)\omega=\varepsilon_I\omega$. Here, as the permittivity is evaluated at the interface the mean value $\varepsilon_I$ must be used. 

Collecting the contributions to the integral equation localized at
the positions of the free charges leaves only polarization terms associated with the surface polarization. Each interface is assumed smooth and is given an orientation so that a
normal vector $\mathbf{n}$ is defined at each of their points. Its
direction is the same as that of the gradient $\nabla\varepsilon$ and therefore
points towards the region of larger permittivity.

After these constructions, the integral equation can be written as:
\begin{equation}
\omega(\mathbf{r})=\frac{1}{\varepsilon_{0}\varepsilon_{I}}\nabla_{r}\varepsilon\cdot\int dV_{r'}\,\nabla_{r}G_{r,r'}\left[\omega(\mathbf{r}')+\frac{\rho(\mathbf{r}')}{\varepsilon_{r'}}\right].\label{eq:IntegralEquation-red}
\end{equation}
Here $\omega$ is the total polarization excluding the contributions
dressing the free charges. The polarization $\omega$ is a volumetric
density localized at a surface so that, in a flat plane, say $z=0$, 
is locally proportional to a delta function $\delta(z)$. Below, by
abuse of notation, the variable $\omega$ will also be used as the
surface charge density that does not include the delta function and,
schematically, an abstract vector on which integral operators act.
The symbol $\gamma$ will be used for the factor that appears at the
interface between media:
\begin{equation}
\gamma=\frac{\Delta\varepsilon}{\varepsilon_{I}}\label{eq:GammaDefinition}
\end{equation}
 where $\varepsilon_{I}$ is the mean value of permittivity at an
interface and $\Delta\varepsilon$ is the difference in relative permittivities. In the general case, each interface can have a different value of this parameter but in the analysis below there is only one interface and a single value. 
The liquid region $\alpha$ is considered to have larger relative
permittivity. The normal therefore points into the region $\alpha$ and $\gamma$ is positive. 

The integral equation can be more succinctly presented as a linear
operator equation of the form
\begin{equation}
\omega=\mathcal{B}\omega+\mathcal{H}\rho.\label{eq:Succict}
\end{equation}
In this expression, $\omega$ is to be understood as a function defined
on the interfaces and excludes any contribution from the polarization
of the free charges. The surface operator $\mathcal{B}$ describes
the interaction between polarization between charges at the boundary
between homogeneous regions. The direct interaction operator $\mathcal{\mathcal{H}}$ 
computes the effect at the surface due to charges in the homogeneous
media. The surface operator integrates the surface density times a
kernel $B(\mathbf{s},\mathbf{s}')$, where $\mathbf{s}$ and $\mathbf{s}'$ 
are three-dimensional coordinates indicating  positions of surface
points. For readability, the shorthand $B_{s,s'}$ is used in several
locations below. Explicitly,
\begin{eqnarray}
\mathcal{B}\omega & = & \int dS_{s'}B(\mathbf{s},\mathbf{s}')\omega(\mathbf{s}')\label{eq:BKernel}\\
B(\mathbf{s},\mathbf{s}') & = & \gamma_{s}\mathbf{n}_{s}\cdot\nabla_{s}G(\mathbf{s},\mathbf{s}')\label{eq:BKernelExpicit}
\end{eqnarray}
Here $dS$ indicates the surface area differential. The explicit form
of the direct polarization operator $\mathcal{H}$ is a volumetric
integration with kernel $H_{s,r}=H(\mathbf{s},\mathbf{r})$, where
the variables used emphasize that the integration is over the charge
distribution in space and the result is observed at a surface point.
This is 
\begin{eqnarray}
\mathcal{H}\rho & = & \int dV_{r'}H(\mathbf{s},\mathbf{r}')\rho(\mathbf{r}')\label{eq:HKernel}\\
H(\mathbf{s},\mathbf{r}') & = & \gamma_{s}\mathbf{n}_{s}\cdot\nabla_{s}G(\mathbf{s},\mathbf{r}')\frac{1}{\varepsilon_{r'}}\label{eq:HKernelExplicit}
\end{eqnarray}
Next, a potential operator $\mathcal{G}$ is defined by integrating
the geometric factor $G_{r,s}=G(\mathbf{r},\mathbf{s})$ against a
surface charge distribution. 
\begin{eqnarray}
\mathcal{G}\omega & = & \int dS_{s}G(\mathbf{r},\mathbf{s})\omega(\mathbf{s}).\label{eq:GKernel}
\end{eqnarray}
The factor $G(\mathbf{r},\mathbf{s})$ also appears in other expressions
in addition to the definition of this operator. The subindices in
those cases refer to the evaluation locations. 

The second term in the Eq. (\ref{eq:EnergyPolarization}) is the excess energy of the system associated with the polarization of the 
interface. This is denoted as $U_p$,  and excludes the direct interactions between bulk charges. Selecting this term only eliminates all singular self-energies of individual ions if they are considered point-like. This is:
\begin{equation}
U_p=\frac{1}{2\varepsilon_{0}}\rho^{T}\mathcal{G}\omega.\label{eq:EnergyPolarization-1}
\end{equation}
Here the transpose symbol $^{T}$ on the charge density indicates
that it is integrated over space against the density $\mathcal{G}\omega$.
Furthermore, the main object of interest is the evaluation of this
expression in the case of a single point-like charge in the bulk.
In that case, $\omega$ is induced by that single charge and the energy
appears as a result of its interaction between the induced charge
and the bulk charge. This is an induced self-energy and its computation
is the main concern of this article. 

\section{Perturbative scheme}\label{section:perturbation}

\subsection{General setup}

Using a perturbative scheme it is possible to express the self-energy of a single ion and the surface polarization caused by it in the  presence of a rough substrate. The scheme expands these quantities as a power series in the amplitude $A$ of the roughness. A dimensionless parameter can be constructed by taking the ratio $A/L$ of the amplitude with respect to the size $L$ of the system. However, the presentation is simplified by working directly with the amplitude.  As the operators used depend on the geometry of the surface,
they also require expansion. The base geometry is the flat case where
all quantities can be exactly evaluated. The results are the closed evaluation
of the first order corrections to the resulting surface density and
self-energy. The calculation is simplified by the fact that that the boundary operator $\mathcal{B}$ is zero in the flat geometry. The integral equation for
the charge density becomes, to first order, a direct evaluation.

The expansion for the polarization $\omega$ is written as
\begin{equation}
\omega=\omega^{(0)}+A\omega^{(1)}+\frac{1}{2}A^{2}\omega^{(2)}\ldots\label{eq:Pert_pmega}
\end{equation}
The superindices in parentheses indicate the order in the expansion.
To first order, the expansion for the operators $\mathcal{H}$, $\mathcal{B}$
and $\mathcal{G}$ read
\begin{equation}
\mathcal{H}=\mathcal{H}^{(0)}+A\mathcal{H}^{(1)}+\ldots,\label{eq:Pert_H}
\end{equation}
\begin{equation}
\mathcal{B}=\mathcal{B}^{(0)}+A\mathcal{B}^{(1)}+\ldots,\label{eq:Pert_B}
\end{equation}
\begin{equation}
\mathcal{G}=\mathcal{G}^{(0)}+A\mathcal{G}^{(1)}+\ldots.\label{eq:Pert_G-1}
\end{equation}
The expansions have a general for but it is already useful to note that in the flat geometry the boundary interaction is zero. This is because it is constructed from the dot product of the electric field between charges in a plane and the normal and these are perpendicular. Namely:
\begin{equation}
\mathcal{B}^{(0)}=\mathcal{B}_{flat}=0.\label{eq:Bflat}
\end{equation}

The self-energy expansion can be written as 
\begin{equation}
U_p=U_p^{(0)}+AU_p^{(1)}+\ldots\label{eq:U1FirstExapnsion}
\end{equation}
with a first order factor
\begin{equation}
U_p^{(1)}=\frac{1}{2\varepsilon_{0}}\rho^{T}(\mathcal{G}^{(1)}\omega^{(0)}+\mathcal{G}^{(0)}\omega^{(1)}).\label{eq:U1inW}
\end{equation}\label{eq:firstU}
The first order correction to the excess self-energy due to the roughness, $\Delta U$, is
\begin{equation}
\Delta U=AU^{(1)}.\label{eq:U1inW2}
\end{equation}

The expansion of these quantities and operators are inserted into the integral equation for the 
Below, the explicit forms of the terms in the operators' expansions
are given. The 0-th order for the equation simply leads to an expression for the 
charge density as  the direct evaluation
\begin{equation}
\omega^{(0)}=\mathcal{H}^{(0)}\rho.\label{eq:U1inW3}
\end{equation}
This expression reproduces the image
method result. The first order equation for the charge density reduces
to 
\begin{equation}
\omega^{(1)}=\mathcal{B}^{(1)}\omega^{(0)}+\mathcal{H}^{(1)}\rho=\mathcal{B}^{(1)}\mathcal{G}^{(0)}\rho+\mathcal{H}^{(1)}\rho.\label{eq:omega1}
\end{equation}
The evaluation of the self energy is
\begin{equation}
U_p^{(1)}=\frac{\rho^{T}}{2\varepsilon_{0}}[\mathcal{G}^{(1)}\mathcal{H}^{(0)}+\mathcal{G}^{(0)}\mathcal{H}^{(1)}+\mathcal{G}^{(0)}\mathcal{B}^{(1)}\mathcal{H}^{(0)}]\rho.\label{eq:energy2}
\end{equation}
 As can be seen, the expressions for the polarization charge and self-energy are direct evaluations that do not
require any operator inversion.

\subsection{Flat and rough geometries }

\begin{figure}
\includegraphics{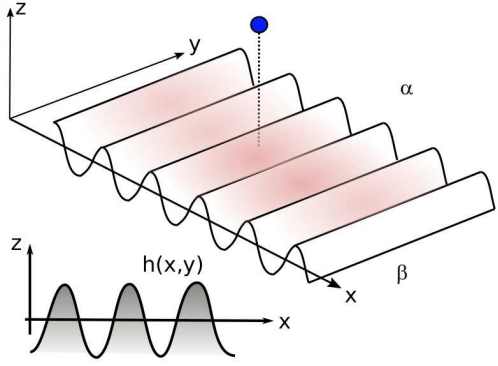}
\caption{\label{fig:Setup}Scheme of the deformed surface. The coordinate system is aligned with
undeformed surface which corresponds to $z=0$. The polarizing ion, shown as a small sphere, 
is in the liquid phase $\alpha$. A sinusoidal deformation is shown
with shape $h(x,y)$. }
\end{figure}

The geometry of a rough substrate can be described as a deformation
of a flat surface. A general roughness pattern can be described
as a Fourier superposition of sinusoidal deformations. To first order
in the deformation, it is then only necessary to evaluate the effect
of such deformations. To this end, we use the coordinate and surface
description indicated in Fig. \ref{fig:Setup}. Cartesian coordinates
are used with generic point $\mathbf{r}=(x,y,z)$, such that the undeformed
surface is located at $z=0.$ Coordinates $(x,y)$ can be used to
parameterize both the flat plane and, by projection, the deformed surface. To make some
expressions below more readable, the coordinates of points
at the surface have a generic three-dimensional coordinate label $\mathbf{s}$. When a point is referred
to by its projection to the plane, the generic two-dimensional vector
$\mathbf{w}=(x,y)$ is used instead. As all calculations refer only
to a single ion; this can be located at a point with coordinates $\mathbf{a}=(0,0,z_{a})$.
The deformation shape is described by a normalized height function $h(x,y)$ and absolute height $Ah$. 
The surface points have deformed surface coordinates $\mathbf{s}$
with the following expansion:
\begin{equation}
\mathbf{s}=\mathbf{s}^{(0)}+A\mathbf{s}^{(1)}+\ldots,\label{eq:sExpansion}
\end{equation}
\begin{equation}
\mathbf{s}^{(0)}=(x,y,0),\label{eq:s0}
\end{equation}
\begin{equation}
\mathbf{s}^{(1)}=(0,0,h(x,y)).\label{eq:s1}
\end{equation}
The main task is to compute the self-energy associated with single
wave shapes. These are taken to oscillate along the x-direction
so that the height is $A_0 \cos(kx+\psi)$ for a specific wavevector value $k$
and phase shift $\psi$. This can be written  
as the real part of a complex exponential with amplitude $A=A_{0}e^{i\psi}$.
The calculation is reduced to consideration of plane waves
of the form $h=\exp(ikx)$. Below, expressions will use both two-
and three-dimensional wavevectors to represent the surface modulation.
The three dimensional wave vector is $\mathbf{k}=(k_{x}=k,k_{y}=0,k_{z}=0)$ 
but the symbol $\mathbf{k}_{2}=(k_{x}=k,k_{y}=0)$ is also used when
calculations are carried out explicitly in two dimensions. As noted
above, a point in the flat plane has three-dimensional coordinate
$\mathbf{s}$ and a two-dimensional representation $\mathbf{w}$.
The single plain wave is then $\exp[ikx]=\exp(i\mathbf{k\cdot s})=\exp(i\mathbf{k}_{2}\cdot\mathbf{w})$.
integrals. 

For a single wave the first order deformation is 
\begin{equation}
\mathbf{s}^{(1)}=(0,0,\exp(ikx))=\exp(i\mathbf{k}\cdot\mathbf{s}^{(0)})\hat{\mathbf{z}}.\label{eq:FirstOrderDeformation}
\end{equation}
The deformed surface has normal $\mathbf{n}$ with perturbative expansion 
\begin{eqnarray}
\mathbf{n} & = & \mathbf{n}^{(0)}+A\mathbf{n}^{(1)}+\ldots\label{eq:normal2-2}\\
\mathbf{n}^{(0)} & = & \hat{\mathbf{z}},\label{eq:normal2-1-1}\\
\mathbf{n}^{(1)} & = & -\exp(i\mathbf{k}\cdot\mathbf{s})\hat{\mathbf{x}}=-i\exp(ikx)\mathbf{\hat{\mathbf{x}}}.\label{eq:normal3-1}
\end{eqnarray}
Here $\hat{\mathbf{x}}$ and $\mathbf{\hat{z}}$ are unit vectors in
the $x$ and $z$ directions. To first order the area differential
of the deformed surface remains $dA=d^{2}\mathbf{w}=dx\,dy$. 

\subsection{Operator perturbations}

The kernels of the operators $\mathcal{B}$, $\mathcal{H}$ and 
$\mathcal{G}$ have general form as indicated in Eqs. \ref{eq:BKernel},
\ref{eq:HKernel}, and \ref{eq:GKernel}. In the perturbative scheme
their 0-th order terms correspond to
their values in the flat case. These are:
\begin{equation}
B_{r,r'}^{(0)}=0,\label{eq:KerB0flat}
\end{equation}
\begin{equation}
H_{s,r}^{(0)}=\frac{1}{\varepsilon_{\alpha}}\gamma\frac{1}{4\pi}\frac{(\mathbf{a}-\mathbf{s})\cdot\mathbf{\mathbf{\hat{z}}}}{|\mathbf{a}-\mathbf{s}|^{3/2}}=\frac{\gamma}{4\pi\varepsilon_{\alpha}}\frac{z_{a}}{(z_{a}^{2}+x^{2}+y^{2})^{3/2}},\label{eq:KetH0Flat}
\end{equation}
and 
\begin{equation}
G_{s,r}^{(0)}=\frac{1}{4\pi}\frac{1}{|\mathbf{a}-\mathbf{s}|}=\frac{1}{4\pi}\frac{1}{(z_{a}^{2}+x^{2}+y^{2})^{1/2}}.\label{eq:KerG0Flat}
\end{equation}

Next, the first-order terms in the operators' expansions are obtained
by determining the changes in their kernels due to the surface deformation.
The deformation changes the relative distance between points and,
in the case of the boundary interaction and direct interaction operators
the relative direction of the normal. In the case of the geometric
Greens function, the change is only due to the modified position of
the surface. This is calculated  using an operator
of the form $\mathbf{s}^{(1)}\cdot\nabla$. In the expressions below,
this operator acts only on the Greens function and its derivatives.
The deformation of the normal vector is considered separately. 

The operator $B_{s,s;}$ is deformed at both positions $s$ and $s'$,
giving rise to terms $B^{(1a)}$, and $B^{(1b)}$. It also receives
a contribution from the change of normal at $w$, namely $B^{(1c)}$. 
\begin{eqnarray}
B_{s,s'}^{(1)} & = & B^{(1a)}+B^{(1b)}+B^{(1c)},\label{eq:B1Split-1}\\
B_{s,s'}^{(1a)} & = & (\mathbf{s}^{(1)}\cdot\nabla_{s})(\mathbf{n}^{(0)}\cdot\nabla_{s})G_{s,s'},\label{eq:B1A}\\
B_{s,s'}^{(1b)} & = & (\mathbf{s}^{(1)}\cdot\nabla_{s'})(\mathbf{n}^{(0)}\cdot\nabla_{s})G_{s,s'},\label{eq:B1B}\\
B_{s,s'}^{(1c)} & = & \mathbf{n}^{(1)}\cdot\nabla_{s'}G_{s,s'}.\label{eq:B1C}
\end{eqnarray}
Explicitly, these are 
\begin{eqnarray}
B_{s,s'}^{(1a)} & = & \frac{\gamma}{4\pi}\frac{\exp(i\mathbf{k}\cdot\mathbf{s}')}{|\mathbf{s}-\mathbf{s'}|^{3}},\label{eq:B1a}\\
B_{s,s'}^{(1b)} & = & \frac{(-\gamma)}{4\pi}\frac{\exp(i\mathbf{k}\cdot\mathbf{s})}{|\mathbf{s}-\mathbf{s}'|^{3}},\label{eq:B1b}\\
B_{s,s'}^{(1c)} & = & \frac{\gamma}{4\pi}\frac{\mathbf{k}\cdot(\mathbf{s}-\mathbf{s}')\exp(i\mathbf{k}\cdot\mathbf{s})}{|\mathbf{s}-\mathbf{s}'|^{3}}.\label{eq:B1c}
\end{eqnarray}

The first-order term for the direct interaction operator has contributions
due to the change in position $H^{(1a)}$ and the change in normal
orientation $H^{(1b)}$. 
\begin{eqnarray}
H_{s,r}^{(1)} & = & H_{s,r}^{(1a)}+H_{s,r}^{(1b)},\label{eq:H1a0-1-2}\\
H_{s,r}^{(1a)} & = & \frac{1}{\varepsilon_{r}}\gamma(\mathbf{s}^{(1)}\cdot\nabla)_{s}(\mathbf{n}^{(0)}\cdot\nabla)_{s}G_{w,r},\label{eq:H1a0-4}\\
H_{s,r}^{(1b)} & = & \frac{1}{\varepsilon_{\alpha}}\gamma(\mathbf{n}^{(1)}\cdot\nabla)_{s}G_{s,r}.\label{eq:H1b0-3}
\end{eqnarray}
Explicitly,
\begin{eqnarray}
H_{s,r}^{(1a)} & = & \frac{1}{\varepsilon_{\alpha}}\frac{\gamma}{4\pi}\frac{e^{i\mathbf{k}\cdot\mathbf{s}}}{|\mathbf{s}-\mathbf{a}|^{3}}\left[\frac{3(\mathbf{\hat{z}}\cdot\mathbf{\mathbf{a}})^{2}}{|\mathbf{s}-\mathbf{a}|^{2}}-1\right]\label{eq:H1a0-2-1}\\
H_{s,r}^{(1b)} & = & \frac{1}{\varepsilon_{\alpha}}\frac{\gamma}{4\pi}\frac{e^{i\mathbf{k}\cdot\mathbf{s}}(\mathbf{s}-\mathbf{a})\cdot\mathbf{k}}{|\mathbf{s}-\mathbf{a}|^{3}}.\label{eq:H1b0-1-1}
\end{eqnarray}
Finally, the geometric Green function has first order expansion of
the form:
\begin{eqnarray}
G_{r,s}^{(1)} & = & (\mathbf{s}^{(1)}\cdot\nabla_{s})G_{rs},\label{eq:G1abs}
\end{eqnarray}
\begin{equation}
G_{r,s}^{(1)}=\frac{1}{4\pi}\frac{e^{i\mathbf{k}\cdot\mathbf{s}}\mathbf{\mathbf{a}\cdot\hat{\mathbf{z}}}}{|\mathbf{a}-\mathbf{s}|^{3}}.\label{eq:G1exp}
\end{equation}

With these explicit expressions the self-energy and surface
polarization density for a single bulk ion are obtained from equations 
\ref{eq:omega1}, \ref{eq:energy2} .
The first order term for the self-energy is
\begin{eqnarray}
U_p^{(1)} & = & \frac{1}{2\varepsilon_{0}}\rho^{T}[\mathcal{G}^{(1)}\mathcal{H}^{(0)}+\mathcal{G}^{(0)}\mathcal{H}^{(1a)}+\ldots\nonumber \\
 &  & \mathcal{G}^{(0)}\mathcal{H}^{(1b)}+\mathcal{G}^{(0)}\mathcal{B}^{(1a)}\mathcal{H}^{(0)}+\ldots\nonumber \\
 &  & \mathcal{G}^{(0)}\mathcal{B}^{(1b)}\mathcal{H}^{(0)}+\mathcal{G}^{(0)}\mathcal{B}^{(1c)}\mathcal{H}^{(0)}]\rho\label{DeltaU1Explicit}
\end{eqnarray}
For readability the expression can be presented as 
\begin{eqnarray}
U_p^{(1)} & = & \frac{q^{2}}{2\varepsilon_{\alpha}\varepsilon_{0}}[\gamma I_{1}+\gamma I_{2a}+\gamma I_{2b}+\ldots\nonumber \\
 &  & \gamma^{2}I_{3a}+\gamma^{2}I_{3b}+\gamma^{2}I_{3c}],\label{eq:U1InGamma}
\end{eqnarray}
 where the $I$ quantities are proportional to the respective terms
in the previous equation. 
\begin{eqnarray}
\omega^{(1)} & = & [\mathcal{H}^{(1a)}+\mathcal{H}^{(1b)}+\mathcal{B}^{(1a)}\mathcal{H}^{(0)}+\ldots\nonumber \\
 &  & +\mathcal{B}^{(1b)}\mathcal{H}^{(0)}+\mathcal{B}^{(1c)}\mathcal{H}^{(0)}]\rho.\label{eq:Omega1Master}
\end{eqnarray}
In a more streamlined way the result can be presented as 
\begin{equation}
\omega^{(1)}=q[\gamma W_{1a}+\gamma W_{1b}+\gamma^{2}W_{2a}+\gamma^{2}W_{2b}+\gamma^{2}W_{2c}],\label{eq:Polarization1W}
\end{equation}
 with the respective identification of terms in the last two equations.

The expressions for the  $I$ and $W$ terms for these last expressions are 
obtained by surface integrations of the respective
integral kernels. Most of the integrations are relatively straightforward,
or at least accessible to symbolic manipulations. However, a couple
of them require precise rearrangements to cancel singularities in
Fourier space, and the use of elliptical coordinates. Therefore, explicit
calculations are provided in the appendices. One of the integral expressions
for $\omega^{(1)}$ does not appear to be expressible as an analytical
form but can be reduced to a  quadrature,

\section{Results }\label{section:results}

\subsection{Flat case}

The scheme we use is based on perturbations of the flat system. In
the description used, the results for that case are of course identical
to the ones obtained by the image method. For an ion of charge $q=Ze$
we have a polarization surface charge
\begin{equation}
\omega_{w}^{(0)}=q\gamma H_{w,s}^{(0)}=q\frac{\gamma}{4\pi\varepsilon_{\alpha}}\frac{\mathbf{a}\cdot\mathbf{\hat{z}}}{|\mathbf{a-\mathbf{s}}|^{3}}.\label{eq:OmegaZero}
\end{equation}
The self-energy associated with the surface charge is then
\begin{equation}
U_p^{(0)}=\frac{1}{8\pi\varepsilon_{0}}\frac{q^{2}}{\varepsilon_{\alpha}}\gamma\frac{1}{4z_{a}},\label{eq:FlatImage}
\end{equation}
which is the standard value obtained by the image method. Namely,
it reflects the interaction of the charge $q$ with its image $q'=\Delta\varepsilon/(2\varepsilon_{I})q=\gamma q/2$,
located at a distance $2z_{a}$ from the source. In the context of applications to confined aqueous electrolytes it is useful to note that this energy is always positive, independent of sign charge. 
This is the case as the liquid phase has larger permittivity than
the bounding material so that $\gamma>0$. Dividing by the thermal
factor $kT$ the result can also be expressed as 
\begin{equation}
\frac{U_p}{k_BT}=\frac{\ell_B}{8}\frac{\gamma Z^{2}}{z_{a}}.\label{eq:FlatImageREduced}
\end{equation}

\subsection{Self-energy for deformed surface }

In the case of the sinusoidal deformation, the self-energy of the
ion due to its interaction with the surface is calculated from expression
Eq. \ref{eq:U1InGamma}. Remarkably, not only the expression reduces,
as noted above, to a direct evaluation, but the required integrals
can be evaluated analytically. Using the evaluations obtained in Appendix A, the final
and central result of this article is that the excess self-energy for an ion is
\begin{eqnarray}
\Delta U_p&=& A\exp^{ikx}  U^{(1)}(k, z)\label{eq:Energy1FullFar}\\
 U^{(1)} & = &\frac{q^{2}}{2\varepsilon_{\alpha}\varepsilon_{0}}\frac{1}{4\pi z^{2}} 
  [\gamma\frac{kz}{4}K_{1}(kz)+\ldots.\nonumber \\
 &  & (-1)\gamma^{2}\frac{(kz)^{2}}{16}K_{0}(kz)]\label{eq:MainResult}
%\Delta U_p&=&A \cos(k x) U^(1)(k, z)\label{eq:Energy1FullFar}\\
% U^{(1) & = &\frac{q^{2}}{2\varepsilon_{\alpha}\varepsilon_{0}}\frac{1}{4\pi z^{2}} 
%  [\gamma\frac{kz}{4}K_{1}(kz)+\ldots.\nonumber \\
% &  & (-1)\gamma^{2}\frac{(kz)^{2}}{16}K_{0}(kz)]\label{eq:MainResult}
\end{eqnarray}
Here $z$ is the elevation of the ion above the mean surface; the subscript (in $z_a$) is omitted. The expression uses the modified Bessel functions of 0-th and 1-st
order, $K_{0}$ and $K_{1}$ respectively. 

The form of the final result indicates that there are two distinct effects created by the surface deformation. The parameter $\gamma$ is in practical applications close to a numerical value of $1$, but the terms proportional to $\gamma$ and $\gamma^2$ have different functional behavior. The powers of $\gamma$ appear in the calculation according to the  number of surface integrations. The term proportional to $\gamma$ is associated with the interaction of the ion with the direct polarization effect it creates, just as in the classic image case, though accounting for the changes of the relative distance to the wall. The term proportional to $\gamma^2$ is associated with the interaction of the ion with the secondary polarization created on the deformed surface due to the direct polarization. This contribution does not appear in the flat case. Evaluation of the term makes clear that it is negative and dominant at high wave-vectors. It is therefore clearly associated with the local curvature of the surface. 

Several aspects of the expression obtained require further discussion.
First, we note that while the perturbative description is standard,
it is not possible to establish its convergence rigorously. Furthermore,
there is an important limitation in the description of the self-energy
at close distances from the surface, as the region where the ion position
$z_{a}$ is within the deformation amplitude $A$, the result is not
well defined. A way to recover sensible results in this region is
discussed in the next section. 

To make the expression obtained more user-friendly, we can consider the region not in the immediate vicinity of the deformation but still within a range where its effects can be felt. For this, we can use the asymptotic form of the modified Bessel functions $K_\nu(z)$. For all index $\nu$ this is $K_\nu(z)\approx [\pi/(2z)]^{1/2}\exp(-z)$. It is also convenient to measure the energy in thermal units $k_{B}T$ a. The real form of the deformation $A\cos(kx+\psi)$ is directly constructed from the previous result just taking a real part of the complex amplitude result. With these approximations and notation, we can write 
\begin{eqnarray}
\frac{\Delta U_p}{k_BT}&=&\frac{A}{\ell_B} \cos(k x+\psi) \frac{U^{(1)}(k, z)}{k_BT}\label{eq:Energy1FullFar2}\\
 \frac{U^{(1)}}{k_BT} & = &\frac{Z^{2}\ell_B}{z^2}\frac{\pi^{1/2}}{128^{1/2}}
  [\gamma (kz)^{1/2}-\gamma^{2}\frac{(kz)^{3/2}}{4}]e^{-k z}\label{eq:MainResultRead}
%\Delta U_p&=&A \cos(k x) U^(1)(k, z)\label{eq:Energy1FullFar}\\
% U^{(1) & = &\frac{q^{2}}{2\varepsilon_{\alpha}\varepsilon_{0}}\frac{1}{4\pi z^{2}} 
%  [\gamma\frac{kz}{4}K_{1}(kz)+\ldots.\nonumber \\
% &  & (-1)\gamma^{2}\frac{(kz)^{2}}{16}K_{0}(kz)]\label{eq:MainResult}
\end{eqnarray}
with the numerical constant $C=(pi/128)^{1/2}$. This expression indicates that the effect of the deformation pattern on the self-energy decays into the bulk of the liquid region with a characteristic length equal to the wavelength of the deformation. The effect is further diminished by the inverse powers of the distance $z$ to the surface. The effect of long wavelength deformations is eliminated by these powers.

%The amplitude is the real part of a complex amplitude and carries
%information about the relative position of the crests of the sinusoidal
%deformation with respect to the projection of the ion location on
%the plane. In the system used the projection is always at the origin.
%Taking $A=A_{0}\exp(i\phi)$ with $A_{0}$real describes a deformation
%shifted away from the origin and the self energy correction is multiplied
%by the real factor $\cos\phi$. Therefore, the expression given is
%the potential above a crest. The correction at any other point has,
%simply, an oscillating behavior with respect to the relative transversal
%position of crests and ions with identical wavelength as the deformation. 

\begin{figure}
\includegraphics{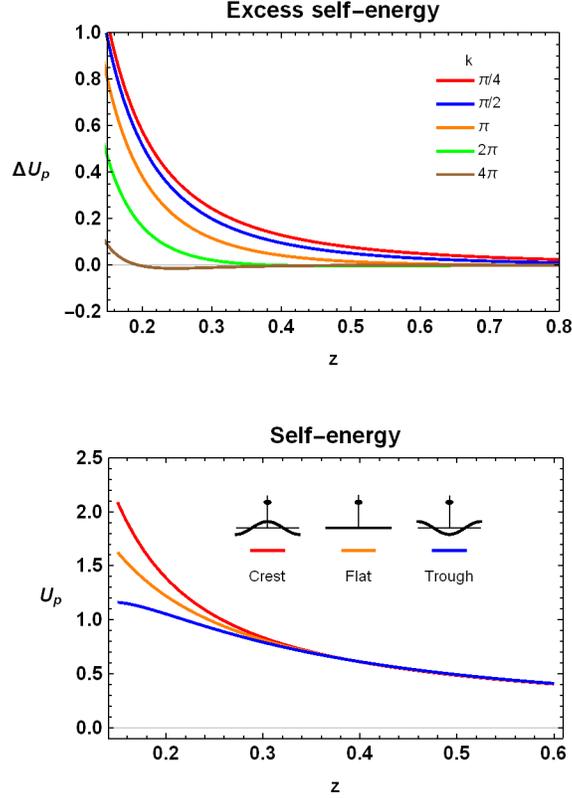}
\caption{\label{fig:potenital-deformation}At the top, the excess self-energy $Delta U_0= A U_p^{(1)})$ for a single ion as a function of height away from the undeformed surface in the presence
of a sinusoidal deformation of amplitude 0.1$\ell_{B}$. The ion is located above a crest.  At the bottom,
the total self-energy $U_p^{(0)}+AU_p^{(1)}$ due to surface polarization
for the ion in the case of a flat surface, and for positions above
a peak and a trough of a surface with sinusoidal deformation with
amplitude 0.1$\ell_{B}$ and wavelength $\ell_{B}$. The energies
are shown in units of thermal energy and all lengths are measured
as multiples of the Bjerrum length.  }
\end{figure}

Results for the full form of the self-energy correction can be readily evaluated.  In
Figure \ref{fig:potenital-deformation}, in the top panel, the self-energy
correction is evaluated for different wavevector values and ion locations
above the plane. The figure shows energies measured as multiples of
the thermal energy scale $kT$ while distances are measured as multiples
of Bjerrum lengths, the evaluation corresponds to a location above
a crest. For large wavelengths the correction is positive. At small wavelengths, the curvature contribution
is negative and can dominate the net result for some values of the
ion position. These results agree with the heuristic description of geometric effects 
of section II. The second panel shows the total self-energy due to polarization, including the flat contribution, for three cases. The ion is considered to be right above a crest, above a trough, and above a flat surface. The deformation considered has 
amplitude and wavevector $A=0.1 \ell_{B}$and $k=1/\ell_{B}$.
As can be seen in the example, the correction can be important within
the plotted region, within one Bjerrum length.

\subsection{Polarization for deformed surface}\label{section:polarization}

The polarization surface charge is a function of ion location $\mathbf{a}$ and the observation point at the surface $\mathbf{s}$. In the expressions presented below this variable will stand for the projected location at the surface $(x,y,0)$. Evaluation of the terms for the polarization, as described in the appendix, produces the following
result:
\begin{eqnarray}
\omega^{(1)} & = & \omega_{q}(\mathbf{a},\mathbf{s})+\frac{q}{4\pi\varepsilon_{\alpha}}(\gamma+\frac{1}{2}\gamma^{2})\frac{e^{i\mathbf{k\cdot s}}}{|\mathbf{s}-\mathbf{a}|^{3}}f_{\omega}(\mathbf{a},\mathbf{s})\label{eq:SurfaceChargeResultMain}
\end{eqnarray}
with
\begin{eqnarray}
\omega_{q}(\mathbf{a},\mathbf{s}) & = & -\frac{q\gamma^{2}}{4\varepsilon_{\alpha}}e^{i\mathbf{k\cdot\mathbf{s}}}\int\frac{d^{2}\mathbf{p}}{(2\pi)^{2}}e^{i\mathbf{p}\cdot\mathbf{w}}e^{-z_{a}p}|\mathbf{p}+\mathbf{k}_{2}|\label{eq:Extraomega}
\end{eqnarray}
and 
\begin{equation}
f_{\omega}(\mathbf{a},\mathbf{s})=-1+\frac{3(\hat{\mathbf{z}}\cdot\mathbf{a})^{2}}{|\mathbf{s}-\mathbf{a}|^{2}}+i\mathbf{k}\cdot\mathbf{s}.\label{eq:f1as}
\end{equation}
 The integration defining $\omega_{q}$ is over 
vectors $\mathbf{p}$ in the two-dimensional space reciprocal to the
flat surface. The form of the integral contribution $\omega_{q}$has
a succinct form but, as noted in the appendix, it is best to subtract
the magnitude $k$ from $|\mathbf{p+\mathbf{k}}_{2}|$ for faster
and reliable convergence. 
\begin{figure}
\includegraphics{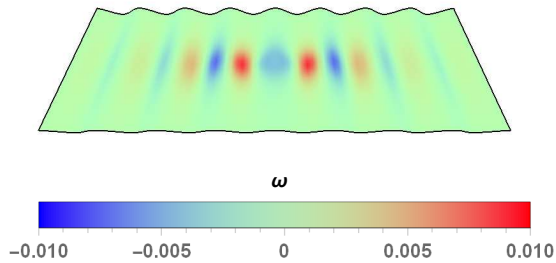}
\caption{\label{fig:surface-charge}Surface charge density for a surface with a single wave deformation
in the presence of a single ion. The ion is located over the central
peak at a distance of $\ell_{B}$ from the undeformed surface. The
wavelength of the deformation is $\ell_{B}$ and its amplitude is
0.1 $\ell_{B}$. The color plot is superimposed on the deformed surface.
Charge density is indicated by color and is presented in units of
$e/\ell_{B}^{2}$. }
\end{figure}

 As with the self-energy and for a single-wave
surface deformations, the dependence on the ion location is periodic
in the direction of the surface oscillation direction. The charge patterns created by
ions over crests and troughs are simply the negative of each other.
The pattern created when the ion over the lines of zero height, is
given by the imaginary part of the expression Eq. \ref{eq:SurfaceChargeResultMain}. 

An example of the excess polarization charge, just due to the deformation,
for a single-wave deformation appears in Figure \ref{fig:surface-charge}.
The figure shows results for $z_{a}=\ell_{B}$, and $k=2\pi/\ell_{B}$.
It can be seen that the pattern decays radially away from the projected
ion location and is coupled to both height and curvature. For the
parameters chosen, the dominant effect is created
by curvature and produces an apparently counterintuitive reduction
of the polarization at the central crest. The crest brings the surface
closer to the ion and thus increases the local polarization. However,
the interaction of the curvature of the deformation with the 0-th
order polarization produces a larger negative contribution. The net charge polarization, which includes the $0-th$ order contribution is still positive everywhere; only the correction from the flat case is negative at that location.

\subsection{Interaction with a dimple}

\begin{figure}
\includegraphics{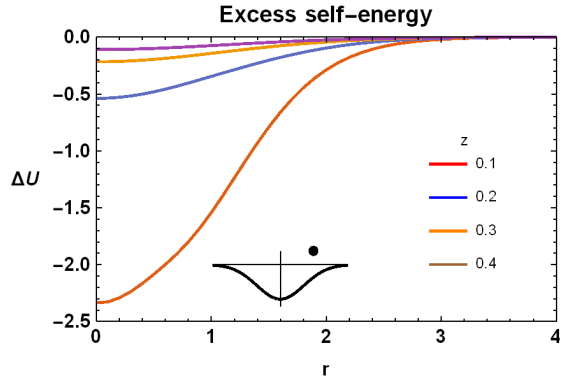}
\caption{\label{fig:dimple}Self-energy for a single ion at different relative radial positions
away from a surface dimple. The different lines correspond to fixed
elevations away from the unperturbed flat surface. The dimple has
a Gaussian shape of depth 0.1$\ell_{B}$ and characteristic width
$\ell_{B}$. The excess energy $U$ is measured in thermal units and
all lengths are measured as multiples of the Bjerrum length. }
\end{figure}

A Fourier superposition allows the determination
of the self-energy of a single ion in the presence of an arbitrary
deformation of the surface. As an example, consider a depression on
the surface, a ``dimple'', with Gaussian form $A\exp[-(x^{2}+y^{2})/2b^{2}]$ 
relative to its center. The relative position of the dimple center and the
ion can be varied to obtain.  Using an amplitude of $-0.1\ell_{B}$ and a width
$b=1\ell_{B}$ produces the results shown in Fig. \ref{fig:dimple} for a set of different
elevations over the flat plane as a function of the radial distance between dimple center and ion. The dimple creates a reduced self-energy
due to the increased distance between ion and surface. 

\section{Results for near approach }\label{section:near}

As noted above, the results for very small distances from the wall
require more careful consideration. The self-energy for point ions
near the flat surface becomes infinite. This is easily regularized
by introduction of finite size ions and is simply avoided by evaluation
away from the interface. However, as the potential is very large near
the surface, its gradients are also large and these have been used
to evaluate the terms in the perturbative expansion. Therefore, this
is not expected to be precise near the deformed surface as is. However,
clear results are still recoverable from the perturbative approach
by proper treatment of the near-surface region. A key result in this
section is the fact that when the set of points at fixed distances from
the deformed surface are considered, the heuristic picture described in section
II still holds and convex regions produce a negative excess self-energy with the opposite result for concave regions.

To construct a more precise solution near the surface is sufficient
to shift the reference plane to the location under consideration. For a given ion location, the reference lane is chosen not as the average undeformed plane but as a plane that exactly matches the surface elevation under the ion. In this way, the ion is at $\mathbf{a}=(0,0,z_a)$ and the deformation height is $h=0$ at the origin. 
Instead of the sinusoidal deformation $h(x)=\exp(ikx)$,
the shifted form is used:
\begin{equation}
h(x)=\exp(ikx)\,-1.\label{eq:ClioseApproach}
\end{equation}
 As written, the deformation consists on the dimpling of the reference
plane; all deformations have a negative (real part) height. The
self-energy for this configuration is then, simply, the same result
as above minus the limit for $k=0$. The result is then
\begin{eqnarray}
\Delta U_p&=&A \cos(k x) U^{(1)}(k, z_a)\label{eq:Energy1FullClose}\\
U^{(1)}(k,z_a)&= & \frac{q^{2}}{2\varepsilon_{\alpha}\varepsilon_{0}}\frac{1}{4\pi z_{a}^{2}}  [\frac{\gamma}{4}\left(kz_{a}K_{1}(kz_{a})-1\right)\nonumber \\
 &  & -\gamma^{2}\frac{(kz_{a})^{2}}{16}K_{0}(kz_{a})]\label{eq:CloseEncounter}
\end{eqnarray}
Similarly, if we consider the close approach to a trough, the height
profile can be taken as $h(x)=A(1-\exp ikx)$ and the free energy
is just the negative of the previous result. Interpolating for all
other positions, or by directly repeating these steps for other locations,
it is possible to conclude that to first-order, the expression above
provides the result for close approach to the surface. These results are consistent with those of the previous section up to differences corresponding to higher orders of the amplitude. Those in the previous section are more useful when exploring the influence of roughness at generic bulk positions. On the other hand, results for close approach allow a clear description of the properties of ions in close contact with the surface. 

\begin{figure}
\includegraphics{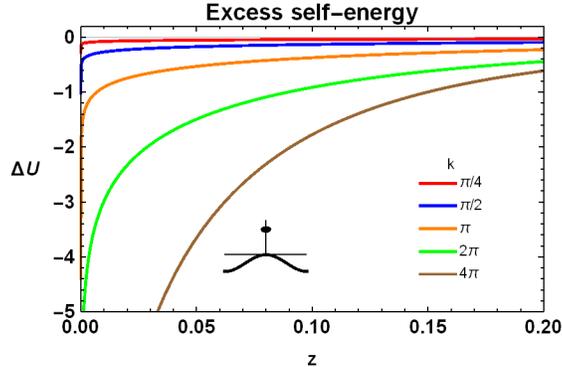}
\caption{\label{fig:nearsurface}Excess self-energy for a single ion as a function of height away from
a sinusoidal peak for different values of the wavevector $k$. The
amplitude of the deformation is 0.1$\ell_{B}$. Distances are measure
The energy is measured in units of thermal energy and all lengths
are measured as multiples of the Bjerrum length. The excess energy
is always negative. }
\end{figure}

The expression obtained is quite satisfying. The self-energy of the flat case diverges
as $1/z_{a}$ but this correction has, upon expansion at short distances,
only a logarithmic $\ln z$ behavior. Namely, the term proportional to $kz(K_1(kz)-1)$ is regular at $z=0$  while the second term is proportional to $K_0(kz) \approx \ln(z)$.  The expression obtained, evaluated near the origin gives the result for a particle near a deformation crest. There, only the term proportional to $\gamma^2$ contributes to the self-energy and its value is always negative in that region. In other words, very near the surface, convexity always reduces the strength of the repulsive interaction. By superposition, the peak of a pimple is less repulsive than a flat surface. The analysis is essentially the same for the bottom of the trough, with a change in sign in the amplitude. There, the repulsive interaction is enhanced and the well of a dimple is more repulsive than the flat surface.   
Evaluation of the self-energy for the crest case is shown in Fig.\ref{fig:nearsurface}. 

\section{Conclusions}\label{section:conclusions}

Possible applications of the results obtained have been outlined in
the introduction section. This section
discusses previously reported results from simulations of systems with deformed
surface geometries, in light of the present work. We indicate possible ways to incorporate the present results into
mean field approaches. 

Recent work \citep{wu2018asymmetric} reported results from
simulations of 1:1 and 2:1 electrolytes confined by a sinusoidally
deformed solid boundary. It was observed that in the absence
of dielectric contrast between liquid and solid, steric effects
reduce the ion population near the surface and that the asymmetric valence case
leads to a net charge distribution near the surface favoring the
monovalent ions. When no dielectric contrast is present, the ion distribution is not noticeably affected
by the shape of the surface; the distributions away from concave and
convex features are nearly identical and their differences are well
within the fluctuation range of the simulation. In contrast, when
the dielectric contrast is turned on, the populations near these features
become clearly different. The concave region exhibits a much stronger
depletion of multivalent ions compared to monovalent ones. Furthermore,
this depletion is stronger at the concave region than at the convex
one. 

The key features observed in the simulation can be the explained based on the results from this article. Namely, turning on the
dielectric contrast between media produces interactions between surface
and ions and mediates interactions between the ions. 
For ions near the surface, the self-energy contribution to the net
electric potential is clearly important as it grows with decreasing
distance to the surface. This contribution depends on the square of
the ions' valencies, which greatly enhances the asymmetry between
species. Both species are repelled from the surface, but the multivalent
ions are more strongly repelled, leading in all cases to a net charging
of the near-surface region. This net charge density has the same sign as the monovalent species.
These observations occur for  both flat and deformed surfaces. Corrections
to this behavior, due to the deformed geometry, consist of enhancement
of ion population near surface peaks and depletion near troughs. These
enhancements and depletions are again larger for the multivalent species.
Thus, the deformation produces an increase in monovalent sign charge
at troughs and a decrease at peaks. As the simulations show, these single 
ion self-energy effects do not disappear at finite particle density. It 
is expected, however, that as concentration increases, these effects would
weaken due to screening from ion-ion interactions. 

The simulations discussed above are computationally expensive
and results that can avoid direct computation are of considerable
value. Several works have so far incorporated consideration of the
self-energy of ions near polarizable surfaces into mean-field approaches
to those systems properties. For electrolytes systems, the Poisson-Boltzmann
description captures the interaction of ions with the mean field created
by surrounding ions, but does not directly include the self-energy
contribution. This can be easily seen as the mean field couples to
the charge of an ion but, in the conditions discussed, the self-energy
is always positive instead of depending on the sign of the valence of the ion.
Instead, an extra contribution must be included in the Boltzmann weight, 
proportional to the exponential
of the self-energy ($U_self$ for a monovalent ion), of the  form $\exp[-Z^{2} U_{self}(\mathbf{x})/kT]$.
Such approach has been implemented in several
previous works where interfaces with dielectric contrast is present,
though with simpler geometries \citep{netz2003variational,Levin2009,buyukdagli2010variational,wang2010fluctuation,wang2013effects}.
The results obtained in this article can be incorporated into these
mean-field schemes to bridge from the single-ion results obtained
here to the case of electrolytes at finite density. As the excess
potential is position dependent these contributions contain information
about the surface shape and how this influences the region near the
surface. When the screening length of the bulk electrolyte
is not too small, the features of the surface are observable in the
particle distribution, as was shown in the simulations discussed above.

\appendix

\section{Evaluation of first order self-energy terms}\label{section:appendix}

The evaluation of the terms that comprise the first order result in
the expansion of the self-energy reduce to integrals over space and
the flat surface. Most of them can be directly evaluated but a few
require limits or coordinate changes that are not trivial. The appendices
present the calculations for all terms, but provide more detail in
the cases where more technical steps are required. 

The self-energy expression is given above in Eq.\ref{eq:U1InGamma}
and repeated here:
\begin{eqnarray}
U_p^{(1)} & = & \frac{q^{2}}{2\varepsilon_{\alpha}\varepsilon_{0}}[\gamma I_{1}+\gamma I_{2a}+\gamma I_{2b}+\ldots\nonumber \\
 &  & \gamma^{2}I_{3a}+\gamma^{2}I_{3b}+\gamma^{2}I_{3c}],\label{eq:U1InGamma-1}
\end{eqnarray}
 The terms $I$ contain, in essence, only geometric quantities. The
results that follow refer to just these expressions. 

The following notation and basic expressions are used. The point ion
location is $\mathbf{a}=(0,0,z_a)$, generic points at the surface are denoted
$\mathbf{s}=(x,y,0)$ and $\mathbf{s}'=(x',y',0)$. These three dimensional
vectors have two dimensional counterparts $\mathbf{w}=(x,y)$ and
$\mathbf{w}'=(x',y')$. The reciprocal vector of a sinusoidal deformation
is $\mathbf{k}$. Its two dimensional version is $\mathbf{k}_{2}=(k_{x},k_{y})=(k,0)$.
All other reciprocal vectors, used in Fourier representations, are
also two dimensional. Most expressions used are derived from two-dimensional
Fourier representations of derivatives of electric potentials and
field evaluated at the surface. For example, the geometric factor
of the normal component $\mathbf{n}\cdot\mathbf{E}$
of the electric field due to a point charge is
\begin{equation}
\frac{1}{4\pi}\frac{\mathbf{\mathbf{z}}\cdot\mathbf{a}}{|\mathbf{a}-\mathbf{s}|^{3}}=\frac{1}{(2\pi)^{2}}\int d^{2}\mathbf{q}\frac{1}{2}e^{-qz_{a}}e^{i\mathbf{q}\cdot\mathbf{w}}\label{eq:BasiFourier2}
\end{equation}
with $q$ the magnitude of the two dimensional reciprocal vector $\mathbf{q}$.
As a preliminary result it is also useful to note that  for two
vectors on the plane, say $\mathbf{w}$, and $\mathbf{w}'$, with
difference $\Delta\mathbf{w}=\mathbf{w}-\mathbf{w}'$, we have
\begin{equation}
\frac{1}{4\pi}\frac{\mathbf{k}\cdot\Delta\mathbf{w}}{|\Delta\mathbf{w}|^{3}}=\lim_{b\rightarrow0}\frac{1}{(2\pi)^{2}}\int d^{2}\mathbf{q}\frac{\mathbf{q}\cdot\mathbf{k}}{2q}e^{-qb}e^{i\mathbf{q}\cdot\Delta\mathbf{w}}.\label{eq:BasicLimt}
\end{equation}
 Use of this limit form facilitates the handling of cancelling divergent contributions in several expressions below.

\subsection{$I_{1}$}

This term is the integration of the geometric factors of $G_{r,s}^{(1)}H_{s,r}^{(0)}.$
\begin{eqnarray}
I_{1} & = & \int dS_{s}\frac{1}{4\pi}\frac{\mathbf{\mathbf{\hat{z}}}\cdot\mathbf{a}\exp(i\mathbf{k}\cdot\mathbf{s})}{|\mathbf{s}-\mathbf{a}|^{3}}\frac{1}{4\pi}\frac{\mathbf{\mathbf{\hat{z}}}\cdot\mathbf{a}}{|\mathbf{s}-\mathbf{a}|^{3}}\nonumber \\
 & = & \frac{1}{(4\pi)^{2}}\int d^{2}\mathbf{w}\frac{z_{a}^{2}\exp i\mathbf{k}_2\cdot\mathbf{w}}{(w^{2}+z_{a}^{2})^{3}}.\nonumber \\
 & = & \frac{1}{(4\pi)}\frac{1}{16z_{a}^{2}}[(kz_{a})^{2}K_{0}(kz_{a})+(kz_{a})K_{1}(kz_{a})].\label{eq:I1Calculation}
\end{eqnarray}
The result is obtained from integration of the angular coordinate
in the $\mathbf{w}$ plane, that gives a Bessel function $J_{0}(kw)$.
The integration against the factor $w/(w^{2}+z_{a}^{2})^{3}$ gives
a result in terms of the Bessel function $K_{2}$, which is then expressed
in terms of $K_{0}$ and $K_{1}$.

\subsection{$I_{2a}$}

This term is the integration of the geometric factors of $G_{r,s}^{(0)}H_{s,r}^{(1a)},.$
\begin{eqnarray}
I_{2a} & = & \int dS_{s}\frac{1}{4\pi}\frac{1}{|\mathbf{s}-\mathbf{a}|}\frac{1}{4\pi}\frac{\exp(i\mathbf{k}\cdot\mathbf{s})}{|\mathbf{s}-\mathbf{a}|^{3}}[\frac{3(\mathbf{\hat{z}\cdot a})^{2}}{|\mathbf{s}-\mathbf{a}|^{2}}-1]\nonumber \\
 & = & \frac{1}{(4\pi)^{2}}\int d^{2}\mathbf{w}\frac{\exp i\mathbf{k}_{2}\cdot\mathbf{w}}{(w^{2}+z_{a}^{2})^{2}}\left[\frac{3z_{a}^{2}}{(w^{2}+z_{a}^{2})}-1\right]\nonumber \\
 & = & \frac{1}{16(4\pi z_{a}^{2})}[3(kz_{a})^{2}K_{0}(kz_{a})+kz_{a}K_{1}(kz_{a})]\label{eq:I2ACalculation}
\end{eqnarray}
The result is obtained from integration of the angular coordinate
in the $\mathbf{w}$ plane resulting on a Bessel function $J_{0}(kr)$.
Then, integration against factors $w/(w^{2}+z_{a}^{2})^{2}$, and
$w/(w^{2}+z_{a}^{2})^{3}$ gives a result in terms of $K_{1}$ and
$K_{2}$. Recursive relations are used to express the result in terms
of $K_{0}$and $K_{1}$.

\subsection{$I_{2b}$}

This term is the integration of the geometric factors of $G_{r,s}^{(0)}H_{s,r}^{(1b)}$.
\begin{eqnarray}
I_{2a} & = & \int dS_{s}\frac{1}{4\pi}\frac{1}{|\mathbf{s}-\mathbf{a}|}\frac{1}{4\pi}\frac{i\mathbf{s}\cdot\mathbf{k}\exp(i\mathbf{k}\cdot\mathbf{s})}{|\mathbf{s}-\mathbf{a}|^{3}}\nonumber \\
 & = & \frac{1}{(4\pi)^{2}}\int d^{2}\mathbf{w}\frac{(i\mathbf{k}_{2}\cdot\mathbf{w})e^{i\mathbf{k}_{2}\cdot\mathbf{w}}}{(w^{2}+z_{a}^{2})^{2}}=\frac{1}{4\pi z_{a}^{2}}\frac{\partial}{\partial k}[\frac{k}{4}K_{1}(kz_{a})]\nonumber \\
 & = & \frac{(-1)}{4\pi z_{a}^{2}}\frac{1}{4}kz_{a}K_{0}(kz_{a}).\label{eq:I2bcalculation}
\end{eqnarray}
The integrand can be expressed as a derivative with respect to $k$
of a term of the form $\exp ikx/(w^{2}+z_{a}^{2})^{2}$. The integral
is a derivative of a multiple of $K_{1}(kz_{a})$. The result follows
from the properties of derivatives of the Bessel function.

\subsection{$I_{3a}$, $I_{3b}$.}

The next two terms are best considered at the same time. These correspond
to the factors $G_{r,s}^{(0)}B_{s,s'}^{(1a)}H_{s,r}^{(0)}$, and $G_{r,s}^{(0)}B_{s,s'}^{(1b)}H_{s',r}^{(0)}$.
\begin{eqnarray}
I_{3\alpha} & = & I_{3a}+I_{3b}\nonumber \\
 & = & \int dS_{s}\int dS_{s'}\left(\frac{1}{4\pi}\right)^{3}\frac{1}{|\mathbf{s}-\mathbf{a}|}\frac{[e^{i\mathbf{k}\cdot\mathbf{s}'}-e^{i\mathbf{k}\cdot\mathbf{s}}]}{|\mathbf{s}-\mathbf{s}'|^{3}}\frac{\mathbf{\mathbf{\hat{z}}}\cdot\mathbf{a}}{|\mathbf{s}'-\mathbf{a}|^{3}}\nonumber \\
 & = & \int\frac{d^{2}\mathbf{w}d^{2}\mathbf{w}'}{(4\pi)^{3}}f_{3\alpha}(\mathbf{w},\mathbf{w}')\label{eq:I3abInW}
\end{eqnarray}
with 
\begin{equation}
f_{3\alpha}(\mathbf{w},\mathbf{w}')=\frac{z_{a}[e^{i\mathbf{k}_{2}\cdot\mathbf{w}'}-e^{i\mathbf{k}_{2}\cdot\mathbf{w}}]}{(z_{a}^{2}+w^{2})^{1/2}|\mathbf{w}-\mathbf{w}'|^{3}(z_{a}^{2}+w'^{2})^{3/2}}\label{eq:Iraaux}
\end{equation}
The integrand has two terms, each the product of three factors with
variables $\mathbf{w}$, $\mathbf{w}'$ and $\mathbf{w}-\mathbf{w}'$.
Each factor can be expressed as a two dimensional Fourier integral.
The integration over the real space coordinates reduces the integral
to the limit of the sum of two terms.
\begin{eqnarray}
I_{3\alpha} & =-\displaystyle{\lim_{b\rightarrow 0}}\frac{1}{8b}\int & \frac{d^{2}\mathbf{Q}}{(2\pi)^{2}}\left[\frac{e^{-bQ}e^{-z_{a}Q}e^{-z_{a}|\mathbf{Q+\mathbf{k}}_{2}|}}{|\mathbf{Q}+\mathbf{k}_{2}|}\right.\nonumber \\
 &  & \left.-\frac{1}{Q}e^{-bQ}e^{-z_{a}Q}e^{-z_{a}|\mathbf{Q-\mathbf{k}}_{2}|}\right]\label{eq:Ialphalim}
\end{eqnarray}
In the second tern of the integral, the $\mathbf{Q}$ variable shifted
to $\mathbf{Q}+\mathbf{k}_{2}$. Then, the limit can be taken to obtain
\begin{eqnarray}
I_{3\alpha} & = & \int\frac{d^{2}\mathbf{Q}}{8(2\pi)^{2}}\left[\frac{Q}{|\mathbf{Q}+\mathbf{k}_{2}|}-1\right]e^{-z_{a}(Q+|\mathbf{Q}+\mathbf{k}_{2}|)}.\label{eq:I3alphaQ2}
\end{eqnarray}
The last expression has two characteristic positions in the $\mathbf{Q}$
plane, namely, the origin and \textbf{$-\mathbf{k}_{2}$}. Factors
appearing in the integrand are functions of distances from these two
positions. For this reason, a change of variables to elliptic coordinates
with loci at the two characteristic locations is natural. These new
variables render an integral with known analytical value. The elliptic
coordinates are
\begin{eqnarray}
\sigma & = & \frac{1}{k}(|\mathbf{Q}+\mathbf{k}_{2}|+Q),\label{eq:Elliptic2}\\
\tau & = & \frac{1}{k}(|\mathbf{Q}+\mathbf{k}_{2}|-Q),\label{eq:Elliptic1}
\end{eqnarray}
The coordinates only cover the upper half plane but by symmetry the
integration over the whole plane is twice the value in the half plane.
The resulting integral is
\begin{equation}
I_{3\alpha}=\int_{-1}^{1}d\tau\int_{1}^{\infty}d\sigma\frac{k^{2}(-\sigma\tau+\tau^{2})e^{-kz_{a}\sigma}}{8(2\pi)^{2}[(\sigma^{2}-1)(1-\tau^{2})]^{1/2}}\label{eq:I3alphaEll}
\end{equation}
The integrals over each of the variables factorize. The integral over
$\tau$ is elementary while the integral over $\sigma$ produces a
modified Bessel function. The net result is 
\begin{equation}
I_{3\alpha}=\frac{1}{(4\pi)}\frac{1}{16z_{a}^{2}}(kz_{a})^{2}K_{0}(kz_{a}).\label{eq:I3alphaC}
\end{equation}

\subsection{$I_{3c}.$}

This integral corresponds to the geometric factors in $G_{r,s}^{(0)}B_{s,s'}^{(1c)}H_{s,r}^{(0)}$.
\begin{eqnarray}
I_{3c} & = & \int dS_{s}\int dS_{s'}\frac{1}{(4\pi)^{3}}\frac{1}{|\mathbf{s}-\mathbf{a}|}\frac{i(\mathbf{s}-\mathbf{s}')\cdot\mathbf{k}e^{i\mathbf{k}\cdot\mathbf{s}}}{|\mathbf{s}-\mathbf{s}'|^{3}}\frac{\mathbf{\mathbf{\hat{z}}}\cdot\mathbf{a}}{|\mathbf{s}'-\mathbf{a}|^{3}}\nonumber \\
 & = & \int\frac{d^{2}\mathbf{w}d^{2}\mathbf{w}'}{(4\pi)^{3}}f_{3c}(\mathbf{w},\mathbf{w}')\label{eq:I3cW}
\end{eqnarray}
with
\begin{equation}
f_{3c}(\mathbf{w},\mathbf{w}')=\frac{i\mathbf{k}_{2}\cdot(\mathbf{w-w}')z_{a}e^{i\mathbf{k}_{2}\cdot\mathbf{w}}}{(z_{a}^{2}+w^{2})^{1/2}|\mathbf{w}-\mathbf{w}'|^{3}(z_{a}^{2}+w'^{2})^{3/2}}\label{eq:I3cWf}
\end{equation}
As in the previous evaluation, the factors can be written as Fourier
integrals and the integration over real space can be carried out.
This leads to a single two dimensional integral in reciprocal space:
\begin{equation}
I_{3c}=\int\frac{d^{2}\mathbf{Q}}{8(2\pi)^{2}}\frac{1}{|\mathbf{Q}+\mathbf{k}_{2}|}\frac{\mathbf{Q}\cdot\mathbf{k}_{2}}{Qk}e^{-z_{a}Q}e^{-z_{a}|\mathbf{Q}+\mathbf{k}_{2}|}.\label{eq:I3cQ}
\end{equation}
Elliptical coordinates can again be used to obtain
\begin{equation}
I_{3c}=\int_{-1}^{1}d\tau\int_{1}^{\infty}d\sigma\frac{k^{2}(-1+\sigma\tau)e^{-kz_{a}\sigma}}{8(2\pi)^{2}[(\sigma^{2}-1)(1-\tau^{2})]^{1/2}}.\label{eq:I3cElliptic}
\end{equation}
The final evaluation produces again a result expressable as a Bessel
function:
\begin{equation}
I_{3c}=-\frac{1}{4\pi z_{a}^{2}}(kz_{a})^{2}K_{0}(z_{a}k).\label{eq:I3cFinal}
\end{equation}

\section{Evaluation of polarization surface charge density. }\label{section:appendixPol}

The first order correction to the surface charge density is expressed
as 
\begin{equation}
\omega^{(1)}=qA[\gamma W_{1a}+\gamma W_{1b}+\gamma^{2}W_{2a}+\gamma^{2}W_{2b}+\gamma^{2}W_{2c}].\label{eq:Omega1App}
\end{equation}
 This appendix describes the evaluation of the terms that appear in
this equation.

\subsection{\textmd{$W_{1a}$}}

This term is the change in the direct polarization contribution due to surface deformation, namely, $H_{s,r}^{(1a)}.$ This is 
\begin{eqnarray}
W_{1a} & = & \frac{1}{4\pi}\frac{\exp[i\mathbf{k\cdot s}]}{|\mathbf{s}-\mathbf{a}|^{3}}\left[-1+\frac{3(\hat{\mathbf{z}}\cdot \mathbf{a})^{2}}{|\mathbf{s}-\mathbf{a}|^{2}}\right]\label{eq:W1a}
\end{eqnarray}

\subsection{\textmd{$W_{1b}$}}

This term is simply the evaluation of the geometric factors in the
change in the direct effect of the point charge on the deformed surface
due to its change in elevation, $H_{s,r}^{(1b)}.$ This is 
\begin{eqnarray}
W_{1b} & =H_{w,a}^{(1)}= & \frac{1}{4\pi}\frac{i\mathbf{k\cdot s}\exp(i\mathbf{k}\cdot\mathbf{s})}{|\mathbf{s}-\mathbf{a}|^{3}}\label{eq:W1b}
\end{eqnarray}

\subsection{\textmd{$W_{2a}$, $W_{2b}$}}

These two terms are again best considered at the same time. They correspond
to the geometric factors in $B_{s,s'}^{(1a)}H_{s',r}^{(0)}$, and
$B_{s,s'}^{(1b)}H_{s',r}^{(0)}$. 
\begin{eqnarray}
W_{2a}+W_{2b} & = & \int dS_{s'}\frac{1}{4\pi}\frac{[e^{i\mathbf{k}\cdot\mathbf{s}'}-e^{i\mathbf{k}\cdot\mathbf{s}}]}{|\mathbf{s}-\mathbf{s}'|^{3}}\frac{1}{4\pi}\frac{\mathbf{\mathbf{\hat{z}}}\cdot\mathbf{a}}{|\mathbf{s}'-\mathbf{a}|^{3}}.\nonumber \\
 & = & \int\frac{d^{2}\mathbf{w}}{(4\pi)^{2}}\frac{z_{a}[e^{i\mathbf{k}_{2}\cdot\mathbf{w}'}-e^{i\mathbf{k}_{2}\cdot\mathbf{w}}]}{|\mathbf{w}-\mathbf{w}'|^{3}\left[z_{a}^{2}+w^{2}\right]^{3/2}}.\label{eq:W2abW}
\end{eqnarray}
The factors in the integrand can be presented as Fourier transforms.
Integration over the surface leads to an integral in reciprocal space:
\begin{eqnarray}
W_{2a}+W_{2b} & = & W_{2\alpha}+W_{2\beta}\nonumber \\
W_{2\alpha} & = & \int\frac{d^{2}\mathbf{Q}}{4(2\pi)^{2}}Qe^{i(\mathbf{Q+\mathbf{k}_{2}})\cdot\mathbf{w}}e^{-z_{a}Q}.\nonumber \\
W_{2\beta} & = & \int\frac{d^{2}\mathbf{Q}}{4(2\pi)^{2}}|\mathbf{Q}+\mathbf{k}_{2}|e^{i(\mathbf{Q+\mathbf{k}_{2}})\cdot\mathbf{w}}e^{-z_{a}Q}.\label{eq:W2abQ}
\end{eqnarray}
To obtain the previous expressions, shifts in the reciprocal variable
similar to those in the computation of the $I_{2}$ terms are employed.
The first integral is proportional to the derivative with respect
to $z_{a}$ of the representation Eq. \ref{eq:BasiFourier2} of the
normal component of the electric field in the plane. This gives:
\begin{equation}
W_{2\alpha}=\frac{1}{4\pi}\frac{1}{2}\frac{\exp[i\mathbf{k\cdot s}]}{|\mathbf{s}-\mathbf{a}|^{3}}\left[-1+\frac{3(\hat{\mathbf{z}}\cdot\mathbf{a})^{2}}{|\mathbf{s}-\mathbf{a}|^{2}}\right]\label{eq:W2alpha}
\end{equation}
where the expression is again presented in terms of the three dimensional
coordinate $\mathbf{s}$.

The second integral $W_{2\beta}$ does not appear to have a closed
form expression. In the main text this expression gives rise to the
term denoted as the remaining quadrature $\omega_{q}$. It is useful
to note that, in numerical evaluations this integral is best computed
using a finite region that is centered at $\mathbf{Q=-k}$. In the
integral the factor $\mathbf{|Q}+\mathbf{k}|$ can be written as the
sum of the terms $(\mathbf{|Q}+\mathbf{k}|-k)$ and $k$. The first
produces an integral that numerically is better behaved while the
second can be analytically evaluated. 

\subsection{$W_{3}$ }

This term contains the geometric factors in $B_{s,s'}^{(1c)}H_{s',r}^{(0)}$.
\begin{eqnarray}
W_{2c} & = & \int dS_{s'}\frac{1}{4\pi}\frac{i(\mathbf{s}-\mathbf{s}')\cdot\mathbf{k}e^{i\mathbf{k}\cdot\mathbf{s}}}{|\mathbf{s}-\mathbf{s}'|^{3}}\frac{1}{4\pi}\frac{\mathbf{\mathbf{\hat{z}}}\cdot\mathbf{a}}{|\mathbf{s}'-\mathbf{a}|^{3}}.\nonumber \\
 & = & \int\frac{d^{2}\mathbf{w}}{(4\pi)^{2}}\frac{i(\mathbf{w}-\mathbf{w}')\cdot\mathbf{k}_{2}z_{a}e^{i\mathbf{k}_{2}\cdot\mathbf{w}}}{|\mathbf{w}-\mathbf{w}'|^{3}(w^{2}+z_{a}^{2})^{3/2}}.\label{eq:W2c}
\end{eqnarray}
In reciprocal space the integral is
\begin{equation}
W_{2c}=\frac{1}{(2\pi)^{2}}\frac{k}{4}e^{i\mathbf{k}_{2}\cdot\mathbf{w}}\int d^{2}\mathbf{Q}\frac{\mathbf{Q}\cdot\mathbf{k}_{2}}{Qk}e^{i\mathbf{Q}\cdot\mathbf{w}}e^{-z_{a}Q}.\label{eq:W2cq}
\end{equation}
 The final result can be obtained by expressing the integrand as a
derivative with respect to $x$ component of $\mathbf{w}$. The result
can be expressed as
\begin{equation}
W_{2c}=\frac{1}{4\pi}\frac{1}{2}\frac{i\mathbf{k}\cdot\mathbf{s}\exp(i\mathbf{k}\cdot\mathbf{s})}{|\mathbf{s}-\mathbf{a}|^{3}}.\label{eq:W2cF}
\end{equation}

%\label{}

\bibliography{solisdeformed}

\end{document}